\documentclass{ws-procs9x6}

\begin{document}
\newcommand{\ea}{\end{equation}}
\newcommand{\ba}{\begin{equation}}
\newcommand{\bev}{\begin{verbatim}}
\newcommand{\beq}{\begin{equation}}
\newcommand{\beqa}{\begin{eqnarray}}
\newcommand{\beqn}{\begin{eqnarray}}
\newcommand{\eeqn}{\end{eqnarray}}
\newcommand{\eeqa}{\end{eqnarray}}
\newcommand{\eeq}{\end{equation}}
\newcommand{\Eev}{\end{verbatim}}
\newcommand{\bec}{\begin{center}}
\newcommand{\eec}{\end{center}}

\title{Solving Renormalization Group Equations \\
by Recursion Relations }
\author{ A. Cafarella, C. Corian\`{o} and M. Guzzi}
\address{Dipartimento di Fisica\\ 
Universita' di Lecce \\
and\\
INFN Sezione di Lecce\\
Via Arnesano 73100, Lecce, Italy \\ 
}

\maketitle

\abstracts{Renormalization Group Equations in integro-differential form 
describing the evolution of 
cascades or resumming logarithmic scaling violations 
have been known in quantum field theory for a long time. 
These equations have been traditionally solved by turning to Mellin moments, 
since in this space they become algebraic.
x-space solutions are less known, but special asymptotic expansions exist 
which allow a fast numerical implementation of these equations. We illustrate 
how the equations can be solved using recursion relations in the 
next-to-leading order approximation.}

\section{Introduction}
\footnote{pr.n.: UNILE-CBR-02-04. Presented at the Intl. Workshop 
``Nonlinear Physics: Theory and Experiment II'', Gallipoli, Lecce, Italy, 
June 28 - July 4, 2002\\
alessandro.cafarella@le.infn.it, claudio.coriano@le.infn.it, marco.guzzi@le.infn.it}
Evolution equations describing the high energy behaviour 
of scattering amplitudes carry significant information on the factorization/ renormalization scale  
dependence of such amplitudes, and allow to link the behaviour 
of processes at a given energy scale to collisions taking place at another (usually much higher) scale. 

In QCD, the accepted theory of the strong 
interactions, confinement forbids the detection of the fundamental 
states of the theory, such as quarks and gluons. However, 
asymptotic freedom allows to separate the perturbative 
dynamics at short distances from the non-perturbative one, due to confinement, through factorization theorems. The functions evolved by these equations are called {\it parton 
distributions}. We will very briefly introduce them below, 
and we will exploit the density matrix formulation of quantum mechanics 
as an analogy to illustrate the matter. 

\section{Nonlocal correlators, Wigner functions and all that}
The mathematical construct which is the closest 
to a parton distribution function (p.d.f.) $q(x, Q)$ 
is a {\it Wigner function}. The analogy is, of course, limited.  

We recall that Wigner's description of quantum mechanics 
via quasi-probabilities of phase space $(\bf{x},p)$
\beq
f(\bf{x},p)= \frac{1}{2 \pi}\int dy \psi^*(x - \frac {\not{h}}{2}y )
\psi(x + \frac {\not{h}}{2}y )
\eeq
is fully equivalent to Schrodinger's formulation \cite{zachos}. 

Differently from Wigner functions, in a p.d.f. a variable $x$, called ``Bjorken x'', takes now the role of the momentum ``p''.
Also, parton distributions are correlation functions of a special type, 
being defined just on the light-cone. In this sense they are not generic 
nonlocal correlators. This limitation, due to the special nature of 
high energy collisions in asymptotically free theories, sets the boundary  
of validity of the parton model approach to QCD.

In QCD one starts by introducing, via arguments based on unitarity, the hadronic tensor, 
which is the key construct describing the collision

\beq
W^{\mu\nu}=\int {d^4 x\over 2 (\pi)^4}e^{i q\cdot x} 
\langle P_A S_A;P_B S_B|\left[J^\mu(0),J^{\nu}(x)\right]| P_A S_A;P_B S_B
\rangle,
\eeq

with $P_A$ and $P_B$ being the momenta of the colliding hadrons and 
$S_A$ and $S_B$ their covariant spins. The $J$'s  are electromagnetic currents. 

The distribution functions that emerge -at leading order- 
from this factorized picture are 
correlation functions of non-local operators in configuration space. 
They are the quark-quark and the quark-antiquark 
correlators.

Their expression simplifies in the axial gauge, in which the 
gauge link is removed by the gauge condition. For instance, the quark-quark correlator takes 
the form 

\beq
\left(\Phi_{a/A}\right)_{\alpha\beta}(P,S,k)=\int {d^4z\over (2 \pi)^4} 
e^{i k\cdot z}
\langle P,S|\overline{\psi}^{(a)}_\beta (0)\psi^{(a)}_\alpha(z)|PS\rangle.
\label{bu}
\eeq
In (\ref{bu}) we have included the quark flavour index $a$ and an index $A$ for the hadron, 
as usual. 
Fields are not time ordered since they can be described by the good 
light cone components of $\psi$ and by $A_T$, a transverse component of the 
gauge field $A_\mu$, as discussed in \cite{RLJ}. 
The non-perturbative information in a collision is carried by 
matrix elements of this type. 

Further considerations allow to show that 
the leading contributions to (\ref{bu}) come from the light-cone region.
The leading expansion of the quark-quark correlator then is of the form

\beq
\int {d\lambda\over 2 \pi}e^{i\lambda x}\langle PS|
\overline{\psi}(0)\psi(\lambda n)|P S\rangle={1\over 2}
\left( \not{p}f_1(x) +\lambda \gamma_5 \not{p} g_1(x) +
\gamma_5 \not{S_T}\not{p}h_1(x)\right), \nonumber \\
\nonumber\\
\label{pdf}
\eeq
where we have used all the four vectors at our disposal 
(spin S, momentum P of the hadron) and introduced invariant amplitudes 
(parton distributions) 
$f_1$, $g_1$, $h_1$, now expressed in terms of a scaling variable x 
(Bjorken x). $n^\mu$ is a light-cone four-vector $(n^2=0)$, approximately 
 perpendicular to the hadron momentum. 

The definition of p.d.f.'s in (\ref{pdf}) involves also an underlying physical scale $Q$ 
($Q >> \Lambda_{QCD}$, with $\Lambda_{QCD}$ being the scale of confinement), not apparent from that equation and characterizing the energy scale 
at which these matrix elements, summarized by (\ref{pdf}), are defined. 
Truly: $f_1=f_1(x,Q^2)$, $h_1=h_1(x,Q^2)$ and so on. 

The role of the 
various renormalization group equations 
is to describe the perturbative change in 
these functions as the scale $Q$ is raised (lowered). Each equation 
involves kernels $(P(x))$ of various types, of well known form, and 
asymptotic expansions of the solutions exist 
(see for instance \cite{FP} and the implementation given in \cite{CS}). Here, 
however, we will illustrate an alternative method to solve these equations 
which is computationally very efficient.  

\section{The equations}
Parton distributions, though fully identified by their operatorial form, 
are not currently calculable from first principles.
In fact, the theory that they describe is a (nonlinear) gauge theory 
characterized by a large QCD coupling constant $\alpha$ at the scale at which they are usually introduced.   
On the other hand, the equations describing their evolution under the renormalization group 
are identified using perturbation theory

\beqa  
 Q^2 {d\over d Q^2} {q_i}^{(-)}(x, Q^2) &=& 
{\alpha(Q^2)\over 2 \pi} P_{(-)}(x, \alpha(Q^2))\otimes q_{i}^{(-)}(x, Q^2)
\nonumber \\
 Q^2 {d\over d Q^2}\chi_i(x,Q^2)
&=& 
{\alpha(Q^2)\over 2 \pi} P_{(-)}(x, \alpha(Q^2))\otimes \chi_i(x,Q^2),
\nonumber \\
\label{quarks}
\eeqa
having defined 
\beq
f\otimes g=\int_x^1\frac{dy}{y} f(\frac{x}{y})g(y)
\eeq

with 
\beq
\chi_i(x,Q^2)={q_i}^{(+)}(x, Q^2) -{1\over n_F} 
q^{(+)}(x, Q^2)
\eeq
for the non-singlet distributions and 
\beqa
&& Q^2 { d\over d Q^2} \left( \begin{array}{c}
q^{(+)}(x, Q^2) \\
G(x, Q^2) \end{array}\right)=
\frac{\alpha}{2 \pi}\left(\begin{array}{cc}
P_{qq}(x,Q^2) & P_{qg}(x,Q^2) \\
P_{gq}(x,Q^2) & P_{gg}(x,Q^2)
\end{array} \right)\otimes 
\left( \begin{array}{c}
q^{(+)}(x, Q^2) \\
G(x,Q^2) \end{array}\right) \nonumber \\
\eeqa
in the singlet sector. Here, $G(x,Q^2)$ is the gluon density, while $q(x,Q^2)$ is 
the quark density. $\alpha$ is the QCD coupling constant.
 
Similar RG equations can be derived for the photon structure 
function $q_\gamma(x, Q^2)$, now with an inhomogeneus term included 
($\alpha_{em}$ is the QED coupling constant)
\beqa  
 Q^2 {d\over d Q^2} {q_\gamma}^{(-)}(x, Q^2) &=& 
{\alpha_{em}\over 2 \pi}\left( 
K^{(0)} + {\alpha\over 2 \pi} K^{(1)}\right)\nonumber \\
&& + {\alpha\over {2 \pi}}\left( P^{(0)}+ {\alpha\over 2 \pi}P^{(1)}
\right)\otimes q_\gamma(x,Q^2). 
\label{photon}
\eeqa
Let's start from the latter equation. The ansatz for the solution of (\ref{photon}) is 
chosen of the form 
\beq
q_\gamma(x, Q^2)={\alpha_{em}\over 2 \pi}
\left(\frac{4\pi}{\beta_0\alpha}\sum_{n=0}^{\infty} \frac{A_n(x)}{n!}\ln^n\left(\frac{\alpha}{\alpha_0}\right)
+ \sum_{n=0}^{\infty} \frac{B_n(x)}{n!}\ln^n\left(\frac{\alpha}{\alpha_0}\right)\right)\nonumber \\
\eeq
and the recursion relations for the functions $A_n(x)$ and $B_n(x)$ appearing 
in the expansion are obtained by comparing terms of the same order in $\alpha^k \log^n(\alpha/\alpha_0)$, with 
k=0,1 and $n=0,1,...$. We use a running QCD coupling at the desired 
perturbative order

\beqa
\frac{d\alpha}{d \log(Q^2)} & = & \beta(\alpha)
 =  -\frac{\beta_0}{4 \pi}\alpha^2 -\frac{\beta_1}{16 \pi^2} \alpha^3
\label{running}
\eeqa
and $\alpha_0\equiv \alpha(Q_0)$, with $Q_0$ being the initial scale at which 
the evolution starts.
In (\ref{running}) $\beta_0$ and $\beta_1$ are the first two coefficients of the QCD beta function.  

The recursion relations are in leading order given by 
\beqa
A_1(x) &=& A_0(x) - K^{(0)}(x) -\frac{2}{\beta_0} P^{(0)}(x)\otimes A_0(x) \nonumber \\
A_{m+1}(x) &=& A_m(x) -\frac{2}{\beta_0} P^{(0)}(x)\otimes A_m(x) \,\,\,\, m=1,2,3... 
\eeqa
and 

\beqa
B_{1}(x) &=& \frac{-\beta_1}{{\beta_0}^2}\left( A_{m+1}(x) - A_m(x) \right) 
- 2 \frac{K^{(1)}(x)}{\beta_0}- \frac{2}{\beta_0} P^{(0)}(x)\otimes B_m(x) -\frac{4}{\beta_0^2}P^{(1)}(x)\otimes A^{(0)}(x) \nonumber \\
B_{m+1}(x) &=& \frac{-\beta_1}{{\beta_0}^2}\left( A_{m+1}(x) - A_m(x) \right) 
-\frac{2}{\beta_0} P^{(0)}(x)\otimes B_m(x) -\frac{4}{\beta_0^2}P^{(1)}(x)\otimes A^{(0)}(x) \nonumber \\
\eeqa
at order $\alpha^2$. The intial condition is easily shown to be of the form 
\beq
\frac{\alpha_{em}}{2 \pi}\left( \frac{4\pi}{\beta_0 \alpha_0}A_0(x) + B_0(x)\right) 
= q_0(x),
\label{initial}
\eeq
with $q_0(x)$ defining the initial functional choice for the parton distribution 
at the lowest scale. 
It is possible to show \cite{storrow} that the freedom in choosing the original 
values for $A_0(x)$ and $B_0(x)$ is not relevant at the order at which we are working 
($\alpha^2$), as far as (\ref{initial}) is satisfied. 

In the case of other equations, such as eqs.~(\ref{quarks}), we get the recursion relations 
\beq
A_{n+1}(x) =  -\frac{2}{\beta_0}P^{(0)}\otimes A_n(x)
\eeq
and
\beqa
B_{n+1}(x) & = & - B_n(x)- \left(\frac{\beta_1}{4 \beta_0} A_{n+1}(x)\right)
- \frac{1}{4 \pi\beta_0}P^{(1)}\otimes A_n(x) -\frac{2}{\beta_0}P^{(0)}
\otimes B_n(x) \nonumber \\
 & = &  - B_n(x) + \left(\frac{\beta_1}{2 \beta_0^2}P^{(0)}\otimes A_n(x)\right)
 \nonumber \\
&&- \frac{1}{4 \pi\beta_0}P^{(1)}\otimes A_n(x) -\frac{2}{\beta_0}P^{(0)}
\otimes B_n(x),  \nonumber \\
\label{recur1}
\eeqa
which are solved with the initial condition $B_0(x)=0$.
The initial conditions for the coefficients $ A_0(x)$ and $B_0(x)$ are specified in a slightly different way from the photon case, with 
$ q(x,Q_0^2) $ now identified as the leading order ansatz for the initial
distribution
\beq
A_0(x)= \delta(1-x)\otimes q(x,Q_0^2)\equiv q_0(x)
\eeq
which also requires $B_0(x)=0$, since we have to 
satisfy the boundary condition 
\beq
A_0(x) + \alpha_0 B_0(x)= q_0(x).
\label{bdry}
\eeq

Again, other boundary choices are possible for $A_0(x)$ and $B_0(x)$ 
as far as (\ref{bdry}) is fullfilled.

\section{Distributional Singularities and Finite Elements}
Once the recursion are given, it remains to be seen how to actually 
implement the method. In practice it is not so easy, but it is easier than 
in other methods \cite{CS}. The codes can range from 1,000 lines for QCD to 
several thousands lines for supersymmetry. The advantage 
of these implementations 
is that the codes can run in few minutes (2-3), compared to the much slower codes 
obtained before. The reason of such improvement is related to the use 
of analytical expressions (in a finite element discretization 
of the integrals) which drastically reduce the computational time required 
to actually perform the recursive integrations. 
There are some important points, however, to keep into account.
The kernels $P(x)$ are defined, in fact, in a  distributional sense and are plagued 
with artificial numerical singularities the most critical ones being tied to  plus (``+'') 
distributions and defined as 
  
\beq
  \int_0^1 dx \frac{f(x)}{(1-x)_+}=\int_0^1 {dy}\frac{f(y)- f(1)}{1-y}.
\eeq
A simple trick to eliminate this singular behaviour is to use the identity  
\beq
 \frac{1}{(1-x)_+}\otimes f(x)\equiv 
\int_x^1\frac{dy}{y}\frac{\,\,y f(y) - x f(x)}{y-x} + f(x)\log(1-x) 
\eeq
and proceed with a finite element discretization of the resulting integral.  

We briefly recall the numerical strategy 
employed in this analysis. 
We define $\bar{P}(x)\equiv x P(x)$ and $\bar{A}(x)\equiv x A(x)$. 
We also define the convolution product  

\beq
J(x)\equiv\int_x^1 \frac{dy}{y}\left(\frac{x}{y}\right) P\left(\frac{x}{y}\right)\bar{A}(y). \ 
\eeq
The integration interval in $y$ at any fixed x-value is partitioned in an array of 
increasing points ordered from left to right 
$\left(x_0,x_1,x_2,...,x_n,x_{n+1}\right)$ 
with $x_0\equiv x$ and $x_{n+1}\equiv 1$ being the upper edge of the integration 
region. One constructs a rescaled array 
$\left(x,x/x_n,...,x/x_2,x/x_1, 1 \right)$. We define 
$s_i\equiv x/x_i$, and $s_{n+1}=x < s_n < s_{n-1}<... s_1 < s_0=1$.
We get 
\beq
J(x)=\sum_{i=0}^N\int_{x_i}^{x_{i+1}}\frac{dy}{y}
\left(\frac{x}{y}\right) P\left(\frac{x}{y}\right)\bar{A}(y) 
\eeq
At this point we introduce the linear interpolation 
\beq
\bar{A}(y)=\left( 1- \frac{y - x_i}{x_{i+1}- x_i}\right)\bar{A}(x_i) + 
\frac{y - x_i}{x_{i+1}-x_i}\bar{A}(x_{i+1})
\label{inter}
\eeq
and perform the integration on each subinterval with a change of variable $y->x/y$ and replace the integral $J(x)$ with 
its discrete approximation $J_N(x)$
to get 
\beqa
J_N(x) &=& \bar{A}(x_0)\frac{1}{1- s_1}\int_{s_1}^1 \frac{dy}{y}P(y)(y - s_1) \nonumber \\
&+& \sum_{i=1}^{N}\bar{A}(x_i) \frac{s_i}{s_i - s_{i+1}}
\int_{s_{i+1}}^{s_i} \frac{dy}{y}P(y)(y - s_{i+1})\nonumber \\
& -& \sum_{i=1}^{N}\bar{A}(x_i) \frac{s_i}{s_{i-1} - s_{i}}
\int_{s_{i}}^{s_{i-1}} \frac{dy}{y}P(y)(y - s_{i-1}). \nonumber \\
\eeqa
Introducing the coefficients  $W(x,x)$ and $W(x_i,x)$, the integral 
is cast in the form 
\beq
J_N(x)=W(x,x) \bar{A}(x) + \sum_{i=1}^{n} W(x_i,x)\bar{A}(x_i)  
\eeq
where
\beqa
W(x,x) &=& \frac{1}{1-s_1} \int_{s_1}^1 \frac{dy}{y}(y- s_1)P(y), \nonumber \\
W(x_i,x) &=& \frac{s_i}{s_i- s_{i+1}}
\int_{s_{i+1}}^{s_i} \frac{dy}{y}\left( y - s_{i+1}\right) P(y) \nonumber \\
& -& \frac{s_i}{s_{i-1} - s_i}\int_{s_i}^{s_{i-1}}\frac{dy}{y}\left(
y - s_{i-1}\right) P(y).\nonumber \\
\eeqa

The results of the integration of the recursion relations 
can be given in analytical form, with obvious care in handling the ``+'' 
distributions. We have solved by this method all the leading twist 
evolution equations of QCD to higher order. Application of the method to 
supersymmetry has also been illustrated \cite{cc}.  
\section{Applications}
As a simple illustration of the method we have included two figures which illustrate 
the variation in shape of the evolved functions with a varying final evolution scale. 
Scaling violations are usually quite small, however they are very important both to uncover 
important new physics and for precision studies. 
Good algorithms are always welcomed. 
\centerline{\bf Acknowledgements}
This work is supported by MIUR and by INFN of Italy (iniz. spec. BA21).  

\begin{figure}[th]
\centerline{\includegraphics[angle=0,width=.9\textwidth]{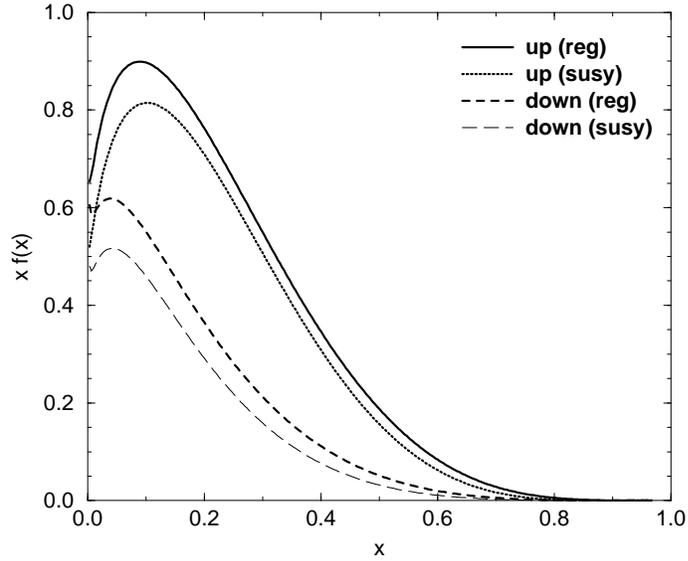}}
\caption{Plot of the supersymmetric evolution of the up and down quarks 
for a final scale of 100 GeV in a specific model. }
\end{figure}
\begin{figure}
\centerline{\includegraphics[angle=0,width=.9\textwidth]{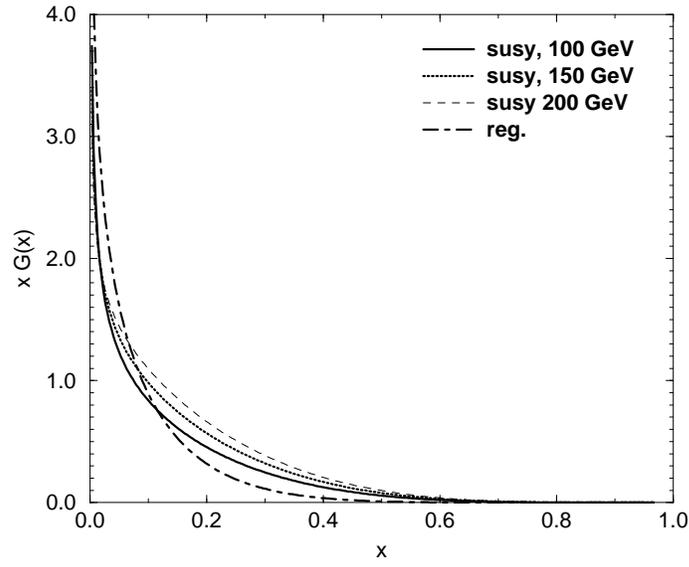}}
\caption{Density of gluons, obtained 
by the recursion method for the QCD (reg.) and Supersymmetric QCD  (SQCD) evolution. 
To vary here is the mass of the supersymmetric partner, the gluino, defining a 
matching scale in the QCD/SQCD evolution.}
\end{figure}


\begin{thebibliography}{0}
\bibitem{CCG} A. Cafarella, C. Corian\'{o} and M. Guzzi, in preparation.
\bibitem{storrow} J.H. Da Luz Vieira and J.K. Storrow, 
Z.Phys. {\bf C51} 24, 1991.  
\bibitem{zachos} T. Curtright and C. Zachos, Prog.Theor.Phys.Suppl. {\bf 135} 244, 1999; J.Phys. {\bf A32} 771, 1999;  C. Zachos, Int.J.Mod.Phys. {\bf A17}, 2002.
\bibitem{FP} W. Furmanski and R. Petronzio, Nucl.Phys.{\bf B195} 237, 1982.
\bibitem{CS} C. Corian\'{o} and C. Savkli, Comput.Phys.Commun.{\bf 118} 236,1999.
\bibitem{cc} C. Corian\'{o}, Nucl.Phys. {\bf B627} 66, 2002; 
 C. Corian\'{o} and A. Faraggi, Phys.Rev. {\bf D65} 075001, 2002. 
\bibitem{RLJ} R.L. Jaffe, Nucl. Phys. B229 (1983) 205.  
\end{thebibliography}
\end{document}